\documentclass[twocolumn,showpacs,preprintnumbers,amsmath,amssymb]{revtex4}

\usepackage{color}

\usepackage{graphicx}
\usepackage{dcolumn}
\usepackage{bm}

\begin{document}
\preprint{APS/123-QED}
\title{Structure properties of ${}^{226}$Th and ${}^{256,258,260}$Fm fission 
fragments:\\ mean field analysis with the Gogny force}
\author{N. Dubray}
 \email{Noel.Dubray@cea.fr}
\author{H. Goutte}
\author{J.-P. Delaroche}

\affiliation{CEA/DAM \^Ile-de-France DPTA/Service de Physique Nucl\'eaire, Bruy\`eres-le-Ch\^atel\\91297 Arpajon cedex, France.}
\date{\today}
\begin{abstract}
The constrained Hartree-Fock-Bogoliubov method is used with the Gogny interaction D1S to calculate potential energy surfaces of fissioning nuclei ${}^{226}$Th and ${}^{256,258,260}$Fm up to very large deformations. The constraints employed are the mass quadrupole and octupole moments. In this subspace of collective coordinates, many scission configurations are identified ranging from symme\-tric to highly asymmetric fragmentations. Corresponding fragment properties at scission are derived yielding fragment deformations, deformation energies, energy partitioning, neutron binding energies at scission, neutron multiplicities, charge polarization and total fragment kinetic energies.
\end{abstract}
\maketitle

\section{Introduction}

Our knowledge of the fission process has made huge progress in recent years with the measurement of mass and charge distributions of fission fragments for 70 fissioning systems~\cite{schmidt00}, performed at the secondary beam facility at GSI. The measured fragment yield distributions have revealed new kinds of systematics on shell structure in nuclear fission, such as transitions from single- and double-humped mass distributions to a triple-humped structure in the vicinity of ${}^{227}$Th. From a theoretical point of view, microscopic self-consistent methods appear to be well suited to study structure effects in fissioning systems, where the sole input is the nucleon-nucleon force. Many studies based on mean field approaches using Gogny or Skyrme forces have recently been devoted to the different fission modes, as for example in ${}^{256-258}$Fm isotopes~\cite{warda02,warda03,warda04,bonneau05a,staszczak05,bonneau06}, where bimodal fission has been experimentally identified~\cite{brandt63,john71,balagna71,flynn72,harbour73,ragaini74,flynn75a,flynn75b,gindler77,hulet80} and analyzed~\cite{moller87,cwiok89}. Furthermore, two-dimensional time-dependent calculations have also been performed for the ${}^{238}$U isotope in the elongation-asymmetry plane, where it appears that fragment mass and total kinetic energy distributions are well reproduced. These calculations have employed the Time-Dependent Generator Coordinate Method treated at the Gaussian Overlap Approximation and used Hartree-Fock-Bogoliubov states~\cite{goutte05}.

The present work, based on the constrained Hartree-Fock-Bogoliubov (HFB) method and the D1S force, is focused on the calculation of structure properties of nascent fission fragments of light and heavy actinides, namely ${}^{226}$Th and ${}^{256,258,260}$Fm. Fragment deformations, deformation energies, energy partitioning, neutron binding energies, neutron multiplicities, charge polarization, and total fragment kinetic energies are calculated for a wide range of fragmentations. This large scale study has been made possible thanks to the new generation of fast computers made available to our laboratory. By the meantime, it is hoped that the calculated structure information here collected for a wide variety of fission fragments will serve as guideline for updating inputs (excitation energy, energy partitioning, neutron binding energy, etc\ldots) to phenomenological evaporation models aimed at calculating prompt neutron emission from, and $\gamma$-ray decay of fission fragments~\cite{lemaire05,lemaire06,kornilov07}.

The paper is organized as follows. In Sec.~II is outlined the constrained HFB method in which, like in Ref.~\cite{goutte05}, quadrupole and octupole mass operators are adopted for external fields. This section also presents the mean field methods used to describe: i) the scission mechanism as well as ii) nascent fission fragments in low energy fission. In Sec.~III results are discussed, among which potential energy landscapes, scission configurations, and fission fragment properties. Fission fragment yields are not considered in this work as they require a dynamical treatment~\cite{goutte05}. Comparisons are made between present predictions and experimental data for total fragment kinetic energy (${}^{226}$Th, ${}^{256}$Fm) and prompt neutron multiplicity (${}^{256}$Fm) of fission fragments.


\section{Self-consistent approach to scission}
\subsection{Constrained Hartree-Fock-Bogoliubov method}
The deformed states of the nuclei under study have been determined using the constrained Hartree-Fock-Bogoliubov (HFB)~\cite{ringetschucka} theory based on the minimization principle of the energy functional, namely
\begin{equation}
\delta <\Phi(\{q_{l0}\})|\hat{H}-\lambda_N\hat{N}-\lambda_Z\hat{Z}-\sum_l\lambda_l\hat{Q}_{l0}|\Phi(\{q_{l0}\})>=0,
\label{eq1}
\end{equation}
where $\hat{H}$ is the nuclear microscopic Hamiltonian, $\hat{Q}_{l0}$ a multipole operator, and $\lambda_N$, $\lambda_Z$, and $\lambda_l$ the Lagrange parameters associated to constraints on nucleon numbers N, Z and average deformations $q_{l0}$, respectively,
\begin{equation}
\begin{array}{cc}
<\Phi(\{q_{l0}\})|\hat{N}|\Phi(\{q_{l0}\})>& =N,\\
<\Phi(\{q_{l0}\})|\hat{Z}|\Phi(\{q_{l0}\})>& =Z,\\
<\Phi(\{q_{l0}\})|\hat{Q}_{l0}|\Phi(\{q_{l0}\})>& =q_{l0},
\end{array}
\label{eq2}
\end{equation}
and where $\hat{Q}_{l0}$ is defined as
\begin{equation}
\hat{Q}_{l0}=(1+\delta_{l,2})\sqrt{\frac{4\pi}{2l+1}}\sum_{i=1}^{A}r_i^l Y_{l0}(\theta_i,\phi_i).
\end{equation}
In the present study, the Hamiltonian $\hat{H}$ is built using the finite range and density-dependent nucleon-nucleon D1S force~\cite{decharge80,berger91}. One-body and two-body corrections for center of mass motion are taken into account in $\hat{H}$. Many calculations have shown that the energy functional derived from this Hamiltonian provides a very satisfactory reproduction of nuclear properties over the whole mass table~\cite{bertsch07} and especially in the actinide region~\cite{delaroche06}. In Eq.~\eqref{eq2} the set of constraints $\{\hat{Q}_{l0}\}$ includes the isoscalar axial dipole, quadrupole, and octupole mass moments $\hat{Q}_{10}$, $\hat{Q}_{20}$ and $\hat{Q}_{30}$, respectively. The dipole moment has been constrained to zero so that the mean position of the nucleus center of mass is located at the origin of the coordinate system. The HFB energy of the deformed system is defined as
\begin{equation}
E_{\text{HFB}}(q_{20},q_{30}) =\, 
<\Phi(q_{20},q_{30})|\hat{H}|\Phi(q_{20},q_{30})>.
\label{ehfb}
\end{equation}
In the present study, the Bogoliubov space has been restricted by enforcing axial symmetry along the $z-$axis and the self-consistent $\hat{T}\hat{\Pi}_2$ symmetry, where $\hat{T}$ is the time-reversal operator and $\hat{\Pi}_2$ the reflection with respect to the xOz plane. The system of Eqs.~(\ref{eq1}) and~(\ref{eq2}) has been solved numerically by iterations for each set of deformations by expanding the single particle states onto axially symmetric harmonic oscillator (HO) bases. For small elongations ($q_{20}$ $<$ $200$~b) a one-center HO basis with $N=14$ major shells has been used while for large elongations ($q_{20}$ $\ge$ $200$~b) a two-centers HO basis with twice $N=11$ major shells has been preferred~\cite{berger85}. The parameters of the one- and two-centers HO bases have been optimized for each set of deformations, and we have checked that the basis sizes are large enough. The potential energies discussed below are the HFB energies defined in Eq.~\eqref{ehfb}.

\subsection{Scission mechanism in the ($q_{20}$, $q_{30}$) plane\label{toposci}}

At large quadrupole moment, it becomes energetically more favorable for a fissioning system to split into two separated fragments, rather than to take on a very elongated shape with a neck. In deformation space, this transition corresponds to an evolution from the so-called fission valley to the so-called fusion valley~\cite{berger84}. If a point A in the fission valley leads to a point B in the fusion valley through a small increment in either one of the deformation parameters, then point A is here defined as a scission point and point B as a post-scission point. Unfortunately, there is no universal way to distinguish a point in the fission valley from a point in the fusion valley, and several criteria have been used in previous studies to achieve this classification. For instance, Bonneau  et al. \cite{bonneau05b,bonneau07} consider that scission occurs when the nuclear interaction between fragments is less than 1~$\%$ of the Coulomb repulsion energy, whereas in refs.~\cite{berger84,goutte05} it was noted that the scission mecanism in ${}^{238}$U and ${}^{240}$Pu is associated to the following three properties: i) the neck between the fragments suddenly vanishes, ii) the hexadecapole moment of the system decreases, and iii) there is a drop in the potential energy of the fissioning system. 
\begin{figure}
\includegraphics[width=8.5cm]{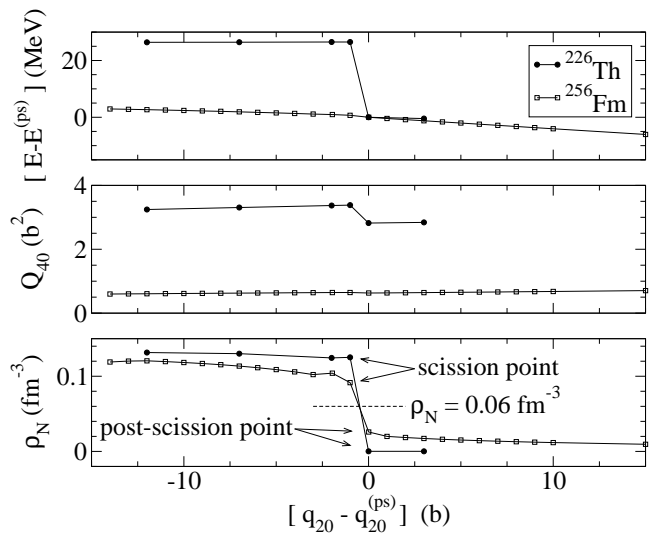}
  \caption{Comparison between symmetric fission of the ${}^{226}$Th and ${}^{256}$Fm nuclei through total energy, hexadecapole moment and minimal density along $z-$axis in the neck $\rho_{\rm N}$. $q_{20}^{\text{(ps)}}$ and $E^{\text{(ps)}}$ represent elongation and HFB energy of the first post-scission (ps) point for each fissioning system, respectively.\label{smooth_scission}}
\end{figure}
Whereas these three criteria appear to be equivalent in the U-Pu region, this is no longer the case for some of the nuclides studied here. As an example, in Fig.~\ref{smooth_scission} we show the evolution of HFB energy, mass hexadecapole moment and minimal density in the neck as functions of the quadrupole moment for the symmetric fission of ${}^{226}$Th and ${}^{256}$Fm. In the top panel, the evolution of the HFB energy shows that scission can either correspond to a sudden loss (${}^{226}$Th curve) or to a smooth decrease (${}^{256}$Fm curve) of the binding energy. The same difference in behaviors can be observed for the evolution of the mean values of the hexadecapole moment $\langle \hat{Q}_{40}\rangle$. These examples clearly illustrate that scission points can be determined neither by an energy- nor an hexadecapole moment-based 
criterion for the scissioning system in ${}^{256}$Fm. The lower panel of Fig.~\ref{smooth_scission} shows that in the symmetric fission of both ${}^{226}$Th and ${}^{256}$Fm, the density in the neck displays two different values before and after scission, with an abrupt drop at scission. In this situation, we define a post-scission configuration as one for which in the matter density along the symmetry axis there is a local minimum that is lower than $\rho=0.06$~fm${}^{-3}$. Using this criterion, we define for each nucleus a set of scission points in the ($q_{20}$, $q_{30}$) plane, that is called the scission line.

Depending on the nucleus and fragmentation, the scission transition is either smooth (e.g. symmetric fragmentation of Fermium isotopes) or abrupt, in the present subspace of collective coordinates. In the first case, the energy of the fissioning nucleus evolves smoothly without discontinuity from outer saddle to scission, becoming the Coulomb repulsion between nascent fragments at large elongation. In the second case, there is an abrupt decrease of energy and hexadecapole moment at scission.

\begin{figure}
\includegraphics[width=8.5cm]{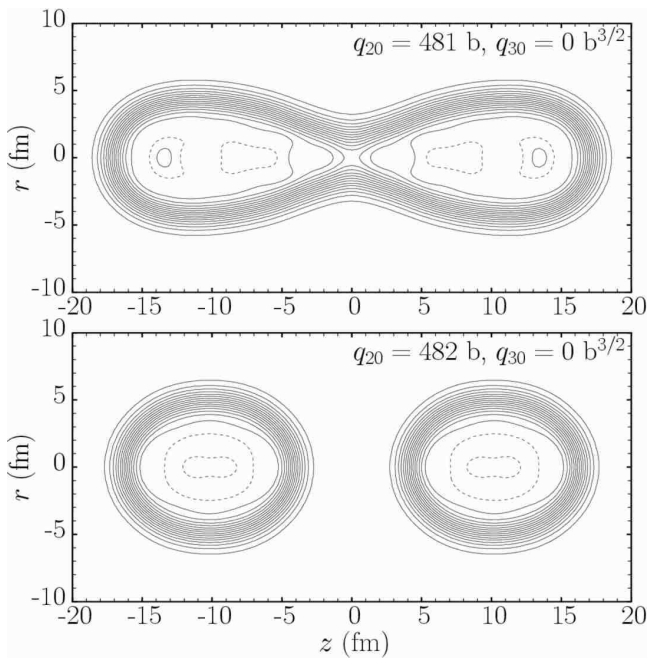}
\caption{Symmetric scission configurations of ${}^{226}$Th in the ($z,r$) space coordinates, before and after scission (upper and lower panels, respectively). The isolines are separated by $0.01$~fm${}^{-3}$. The dashed isoline corresponds to $\rho=0.16$~fm${}^{-3}$.\label{examplescission226th}}
\end{figure}

\begin{figure}
\includegraphics[width=8.5cm]{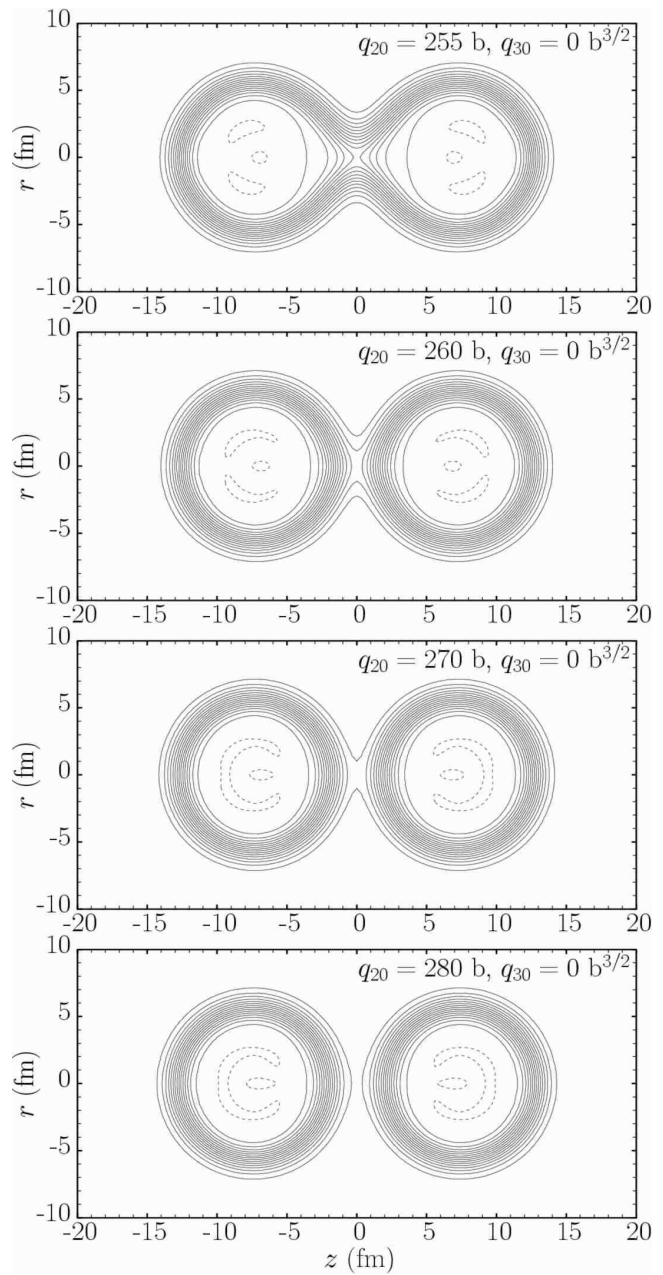}
\caption{Same as Fig.~\ref{examplescission226th} for the symmetric scission of ${}^{256}$Fm.\label{examplescission256fm}}
\end{figure}

Figures \ref{examplescission226th} and \ref{examplescission256fm} show the evolution of the nuclear density at large elongation for the symmetric fragmentation of ${}^{226}$Th and ${}^{256}$Fm, respectively. While ${}^{226}$Th displays a very elongated shape at the scission point (upper panel) and two prolate fragments at the post-scission point (lower panel), ${}^{256}$Fm symmetric fission leads to two nearly spherical fragments separating smoothly. In the literature, these quite different ways of fissioning are called Elongated Fission (EF) and Compact Fission (CF), respectively~\cite{hulet89,moller01}. In the ${}^{256}$Fm symmetric fission case, CF is currently explained by the proximity of double-magicity of the fragments ($Z=50$, $N=82$)~\cite{mustafa78,gonnenwein99}.


\subsection{From scission to fragments}

The main purpose of identifying the scission configurations of the nuclear shape using the method described in the preceding section is to obtain information on fragment properties and distributions. The underlying assumption is that, once a scission configuration is reached, splitting of the nucleus will occur irremediably yielding two separated fragments moving away from each other under the action of their mutual Coulomb repulsion. Observable fragment properties such as kinetic energy or excitation energy can then be inferred from the characteristics of the nascent fragments - as distance between centers of mass, deformations, \ldots - at scission. It is important to stress that the fragment properties derived in such an analysis will not necessarily all correspond to those observed in experiments, since some of the configurations found at scission may not occur with significant probability in the fission process.

For each scission point, a sharp cut is made at the neck position $z_{\rm N}$ on the $z-$axis, which serves to define the light (L) and heavy (H) fragments. Some fragment properties, namely quadrupole and octupole deformations, masses and charges, distances between centers of charge and mass, are next calculated as mean values
\begin{equation}
\langle\hat{O}\rangle_{\rm L}\equiv 2\pi \int_{-\infty}^{z_{\rm N}}dz 
\int_{0}^{\infty}r.dr\, \hat{O}\rho(z,r),
\end{equation}
\begin{equation}
\langle\hat{O}\rangle_{\rm H}\equiv 2\pi \int_{z_{\rm N}}^{\infty}dz 
\int_{0}^{\infty}r.dr\, \hat{O}\rho(z,r),
\end{equation}
where $\rho$ is the nuclear density and $\hat{O}$ a one-body operator.

We have checked that all first multipole moments from $Q_{20}$ to $Q_{60}$ of the fissioning system are continuous along scission lines, which ensures that the scission configurations analyzed form a continuous set from which fragment properties can be consistently derived.
At this stage a few remarks are in order, namely: i) the adopted sharp cut assumption inevitably leads to non-integer values for calculated fragment charges and masses; ii) as our model is restricted to two collective coordinates, only one scission configuration is predicted for any fixed ($q_{2s}$, $q_{3s}$) value. As a consequence, the set of fragment pairs here deduced is only a fraction of all possible pairs which would be formed if the constrained HFB calculations were extended to include other collective coordinates. Nascent fragments associated with different scission configurations may be found having nearly the same proton and neutron numbers. As charge and mass fission fragment yields are outside the scope of the present static model, such fragmentations are considered having the same weight in figures shown below where they will display multiple-values, for example when plotted as a function of fragment mass.

\section{Results}
\subsection{Potential energy landscapes}

\begin{figure*}
  \includegraphics[width=17cm,angle=0]{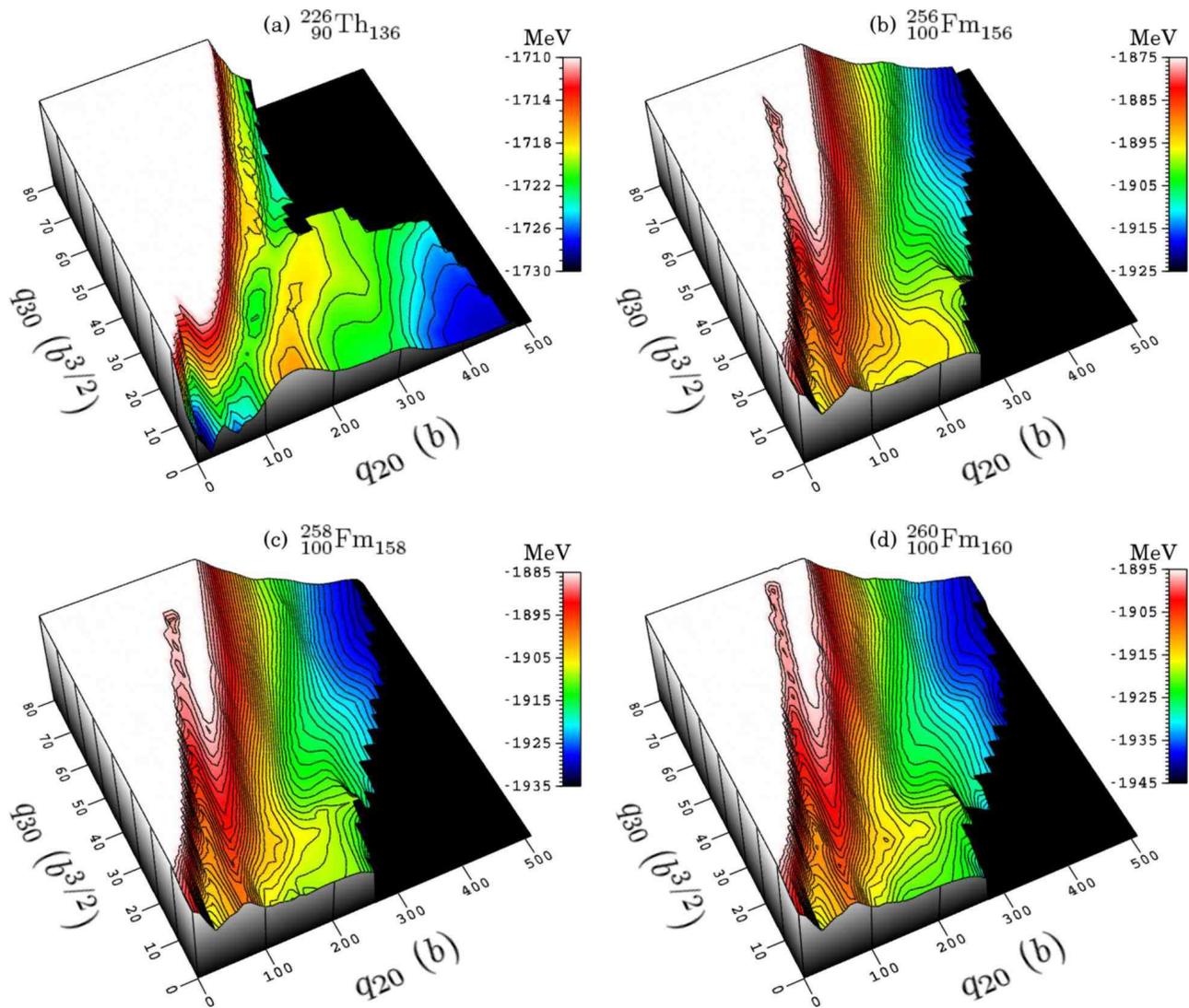}
  \caption{(Color online) Potential energies (MeV) as functions of the $q_{20}$ (b) and $q_{30}$ (b${}^{3/2}$) mass moments for ${}^{226}$Th (a), ${}^{256}$Fm (b), ${}^{258}$Fm (c), and ${}^{260}$Fm (d). Post-scission points are not plotted.\label{allpes}}
\end{figure*}

The potential energies have been calculated on a ($q_{20}$, $q_{30}$) mesh from ($q_{20}=0$, $q_{30}=0$) to ($q_{20}=q_{2s}$, $q_{30}=q_{3s}$), where ($q_{2s}$, $q_{3s}$) belong to the scission lines. With the chosen mesh dimensions $\Delta q_{20}=10$~b and $\Delta q_{30}=4$~b${}^{3/2}$, each of the potential energy landscapes shown in Fig. \ref{allpes} are generated with approximately 600 calculated values. For convenience, the range of potential energies shown is limited to 20 MeV for ${}^{226}$Th (see Fig. \ref{allpes}(a)) and to 50 MeV for the three Fm isotopes (Figs. \ref{allpes}(b), \ref{allpes}(c), \ref{allpes}(d)). Isolines are separated by 1 MeV.

The topological properties displayed by the four landscapes are quite contrasted. We first notice that the lowest potential minima of ${}^{226}$Th and ${}^{256-260}$Fm are all soft against quadrupole and octupole deformations, which should favor coupled quadrupole and octupole vibrations at low excitation energies. These are the common features expected for these nuclides at normal deformations. As axial deformation increases beyond the inner barrier, a well defined superdeformed (SD) potential minimum is taking place only for ${}^{226}$Th. The SD potential minimum is vanishing for ${}^{256-260}$Fm as discussed previously for actinides with neutron number $N>156$ \cite{delaroche06}.

Beyond the SD potential minimum, a valley a few MeV deep is showing up in ${}^{226}$Th for asymmetric deformation all the way to a scission point with large left/right asymmetric fragmentation. An isomeric minimum appears for $q_{20}=140$~b, $q_{30}=20$~b${}^{3/2}$. At elongation $q_{20}>150$~b, a symmetric valley is also observed until the scission point $q_{2s}=500$~b is reached. As scission energies are similar in both valleys, symmetric and asymmetric fission modes are expected to compete in this nucleus. For ${}^{256-260}$Fm, the potential landscapes display similar and smooth patterns beyond the first axial barrier. In contrast to ${}^{226}$Th, we observe that: (i) a shallow asymmetric valley is identified for $q_{30}>30$~b${}^{3/2}$, (ii) the fall-off of the potential landscapes versus elongation for the latter nuclides is smooth for asymmetries $q_{30}> 50$~b${}^{3/2}$, (iii) the scission lines display approximately smooth and linear trajectories over the ($q_{20}$, $q_{30}$) plane, and (iv) a symmetric valley is gradually developing beyond $q_{20}=100$~b as N grows from 156 to 160. This last feature is not inconsistent with the observation of a transition from asymmetric to symmetric mass division in fission, in going from ${}^{256}$Fm to ${}^{258}$Fm \cite{flynn72,hoffman80b,hulet89,hoffman89}. Whether or not this transition can be further analyzed with the present static mean field approach will be discussed below.

\subsection{Scission lines over the ($q_{20}$, $q_{30}$) plane}

\begin{figure}
\includegraphics[width=8.5cm]{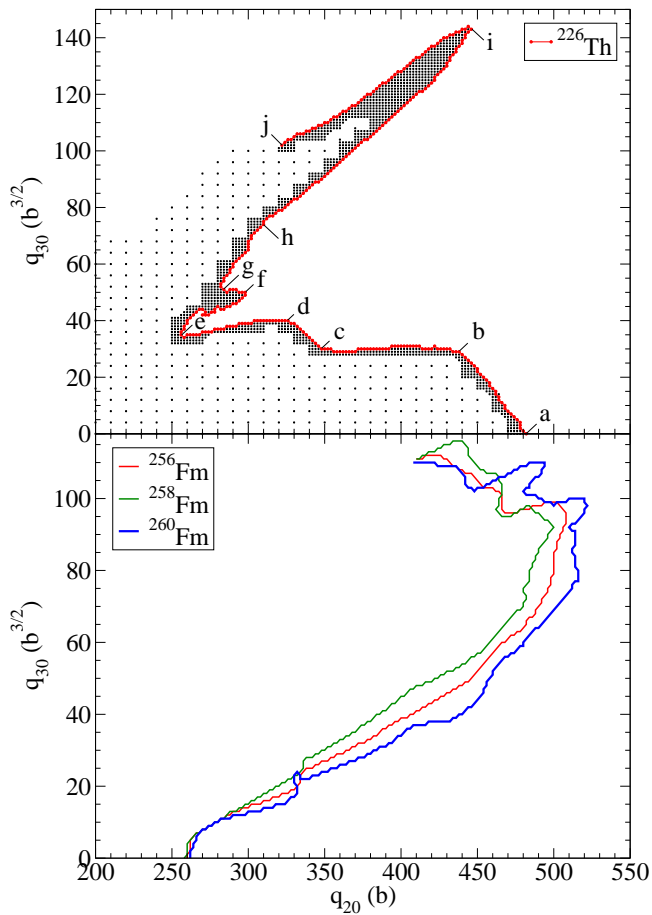}
\caption{(Color online) Upper panel: the scission line for ${}^{226}$Th is shown over the ($q_{20}$, $q_{30}$) plane as a continuous curve in red color along which are marked symbols a, b, c,\ldots j. The black dots, representing single HFB calculations, are shown to illustrate the densening of the mesh used close to the scission line. Lower panel: scission lines for ${}^{256-260}$Fm.\label{scission_lines}}
\end{figure}

The mesh sizes $\Delta q_{20}$ and $\Delta q_{30}$ so far adopted are well suited for performing a survey of potential energy landscape properties. In the vicinity of scission points the step sizes have been dramatically reduced to $\Delta q_{20}=2$~b and $\Delta q_{30}=1$~b${}^{3/2}$, in order to define scission points with high precision. For each nucleus, approximately two to three hundred scission points are used to define a scission line. This is illustrated for ${}^{226}$Th in the upper panel of Fig.~\ref{scission_lines}, where each point in the ($q_{20}$, $q_{30}$) plane corresponds to a single HFB calculation. Only configurations before scission are shown. The curve in red color is the scission line, which is made of all exit points ($q_{2s}$, $q_{3s}$). To ease forthcoming discussions, a few scission points have been labeled with letters a, b, c,\ldots, j.

Most of the constrained HFB calculations at given ($q_{20}$, $q_{30}$) values are performed using as a starting point the generalized density matrix $\mathcal{R}$~\cite{ringetschuckb} obtained at ($q_{20}-\Delta q_{20}$, $q_{30}$). However, in a few cases, it has been necessary to start from the density matrix calculated at either ($q_{20}$, $q_{30}-\Delta q_{30}$) or ($q_{20}+\Delta q_{20}$, $q_{30}$), in order to reach all possible fragmentations. For example, the segments defined between the labels b and c and between the labels d and e were determined increasing asymmetry and decreasing elongation, respectively.

The scission lines determined for ${}^{226}$Th and for ${}^{256}$Fm, ${}^{258}$Fm and ${}^{260}$Fm are shown in the upper and bottom panels of Fig. \ref{scission_lines}, respectively. The lines for the Fm isotopes display similar features. The symmetric scission configurations are found at $q_{2s}=270$~b. Beyond this point, $q_{2s}$ and $q_{3s}$ increase gradually until $q_{2s}$ reaches a maximum for $q_{2s}\simeq 500$~b where asymmetry takes on values in the range $q_{3s}=80-100$~b${}^{3/2}$. For higher asymmetries, the scission lines display wiggling patterns and are quite similar. The trajectory followed by the ${}^{226}$Th scission line over the ($q_{20}$, $q_{30}$) plane is quite different. First, the nucleus stretches and gets an elongation nearly twice as large as the one for Fm nuclides before symmetric scission takes place. Next, elongation decreases as asymmetry increases until $q_{2s}$ reaches a minimum for $q_{2s}=250$~b, (label e in Fig. \ref{scission_lines}). Except for the point on the scission trajectory marked with the label f, $q_{2s}$ and $q_{3s}$ increase smoothly until the point labeled i is reached. Beyond this point located at ($q_{2s}=444$~b, $q_{3s}=142$~b${}^{3/2}$), both $q_{2s}$ and $q_{3s}$ decrease until the scission point labeled j is reached. Although the scission line is defined beyond the point labeled j, this segment lies in a ($q_{20}$, $q_{30}$) region where potential energy is sharply raising. Therefore, the corresponding scission configurations will not be reached in low-energy fission and they will not be considered in the rest of this work. For the same reason, the Fm scission points beyond ($q_{2s}=410$~b, $q_{3s}=111$~b${}^{3/2}$) will also be discarded.

\subsection{Energy along scission line}

\begin{figure}
\includegraphics[width=8.5cm]{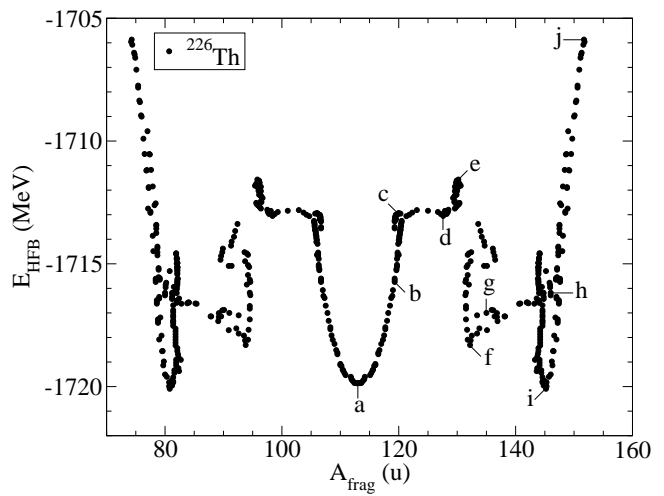}
  \caption{${}^{226}$Th. Potential energy along the scission line as a function of fragment mass. The symbols a, b, c,\ldots j have the same meaning as in Fig.~\ref{scission_lines}. See text for more details.\label{e226Th}}
\end{figure}

\begin{figure}
\includegraphics[width=8.5cm]{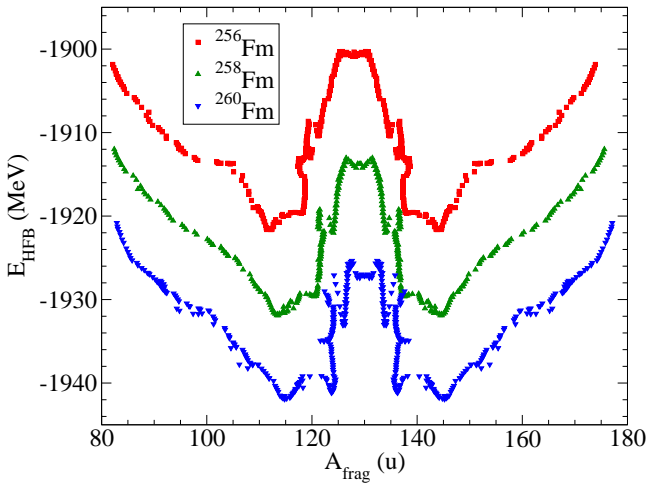}
  \caption{(Color online) Same as Fig.~\ref{e226Th} for ${}^{256-260}$Fm.\label{allfm}}
\end{figure}

The potential energies $E_{\text{HFB}}$ along scission lines are shown for ${}^{226}$Th and ${}^{256-260}$Fm as functions of the fragment mass $A_{\text{frag}}$ in Figs. \ref{e226Th} and \ref{allfm}, respectively. These energies take on identical values on both sides of the symmetric fragmentation where $A_{\text{frag}}=A/2$. On these figures, each solid dot is for a single HFB calculation.

In ${}^{226}$Th one principal and two secondary minima are observed which are likely to represent the most probable fragmentations in low energy fission. Hence, both symmetric ($A_{\text{frag}}\simeq 113$) and two asymmetric modes ($A_{\text{frag}}\simeq 132$ and $A_{\text{frag}}\simeq 145$) are expected for this nucleus. Fragment charges in the symmetric mode ($Z_{\text{frag}}\simeq 45$) and the two asymmetric modes ($Z_{\text{frag}}\simeq 52$ and $Z_{\text{frag}}\simeq 57$) appear in good agreement with those found in the triple-humped mass/charge distribution measured a few years ago and analyzed in terms of the superlong ($Z_{\text{frag}}=45$), standard I ($Z_{\text{frag}}=54$) and standard II ($Z_{\text{frag}}=56$) fission channels \cite{schmidt01b}.

Figs.~\ref{scission_lines} and~\ref{e226Th} show that there is a correlation between the structures in the potential energy along the scission line and the behavior of the scission line in the ($q_{20}$, $q_{30}$) plane. In order to better visualize this correspondence, the potential energy of characteristic scission configurations labeled as a, b, c,\ldots in Fig.~\ref{scission_lines} is displayed in Fig.~\ref{e226Th}. One observes that, as $A_{\text{frag}}$ increases, i) the scission line shifts from symmetric to asymmetric mass division following an irregular trajectory over the ($q_{20}$, $q_{30}$) plane, and ii) to each labeled scission configuration is associated a break in the $E_{\text{HFB}}$ energy values. It thus seems that the competition between symmetric and asymmetric fission of ${}^{226}$Th is tied with the static structure properties of the fissioning system along the scission line. The asymmetric scission configurations calculated for $A_\text{frag}\simeq 132$ and $A_\text{frag}\simeq 145$ coincide with the points marked with the symbols f and i, respectively, in Fig.~\ref{scission_lines}.

The absolute minima in $E_{\text{HFB}}$ for ${}^{256-260}$Fm along the scission line take place for asymmetric fragmentation with $A_{\text{frag}}\simeq 145$, property which correlates rather well with the location of a peak in the fragment mass distributions identified for $A_\text{frag}\simeq 142$ in the ${}^{256}$Fm mass-yield measurements\cite{flynn72}. Symmetric fission is not energetically favored as $E_{\text{HFB}}$ displays a maximum for $A_{\text{frag}}=A/2$, in contrast to the above results for ${}^{226}$Th. However we observe that the difference in energy between the maximum and minimum  values taken by $E_{\text{HFB}}$ for $A_{\text{frag}}=A/2$ and $A_{\text{frag}}\simeq 145$ decreases from 22 MeV to 16 MeV as the mass of Fm isotopes increases from $A=156$ to $A=160$. Although this feature would favor a transition from asymmetric to symmetric fission, making a more definite conclusion on this transition requires a full dynamical calculation in which both potential energy and tensor of inertia from ground state deformation to scission configurations play a role.


\subsection{Fragment deformations}

\begin{figure}
\includegraphics[width=8.5cm]{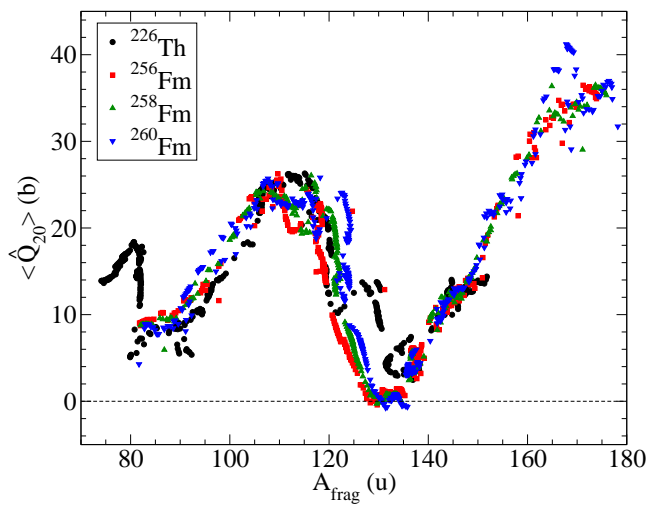}
  \caption{(Color online) Axial mass quadrupole moments $\langle\hat{Q}_{20}\rangle$ of the nascent fission fragments for ${}^{226}$Th and ${}^{256-260}$Fm.\label{fragdefq2}}
\end{figure}

\begin{figure}
\includegraphics[width=8.5cm]{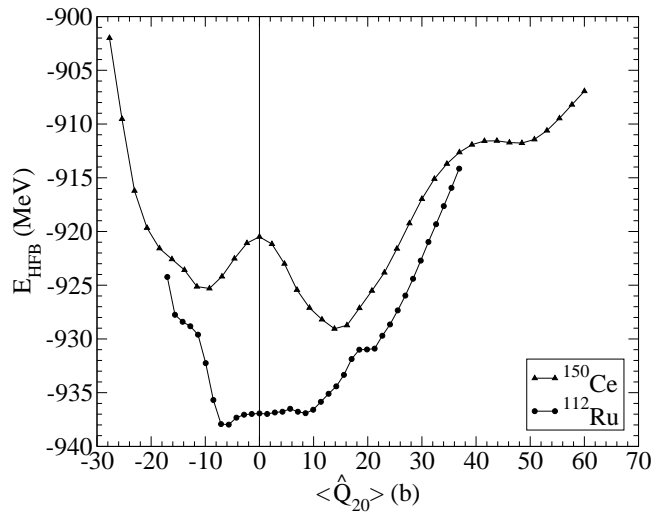}
  \caption{${}^{112}$Ru and ${}^{150}$Ce potential energy curves from constrained HFB calculations    restricted to axially-symmetric and left-right-symmetric shapes as a function of axial quadrupole deformation~\cite{hilaire07}. The potential energy of ${}^{150}$Ce has been arbitrarily increased by 295~MeV to ease comparison between curves.\label{RuCe}}
\end{figure}

The axial mass quadrupole moment of the nascent fission fragments along scission lines is plotted on Fig.~\ref{fragdefq2} for the four studied fissioning systems.
The most striking feature is that the fragment deformations do not significantly depend on the fissioning system. The four curves are almost superimposed and have the expected saw-tooth structure: minima are found for $A\simeq 86$ and $A\simeq 130$, and maxima for $A\simeq 112$ and $A\simeq 170$. Indeed, on the one hand, strong spherical shell effects for $N=80$ and $Z=50$ stabilize spherical fragments of Tin isotopes at scission. In the case of ${}^{226}$Th a minimum with $\langle \hat{Q}_{20}\rangle\simeq 5$~b is also observed, corresponding to Krypton isotopes with $A\simeq 86$. This effect is driven by the neutron magic number $N=50$. On the other hand,  well-deformed Ruthenium isotopes ($Z=44$ and $N\simeq 68$) are here predicted with  $\langle\hat{Q}_{20}\rangle\simeq 22$~b. This deformation corresponds to a shallow secondary minimum of the potential energy curve of the Ruthenium isotopes as a function of quadrupole deformation, as illustrated in Fig.~\ref{RuCe} for ${}^{112}$Ru. Very heavy fragments around $A\simeq 170$ are also predicted to be well-deformed. This shell effect is associated to the deformed magic numbers $Z=58$ and $N=92$ at $\langle\hat{Q}_{20}\rangle\simeq 15$~b. The potential energy curve for ${}^{150}_{\phantom{1}58}$Ce is also plotted in Fig.~\ref{RuCe} as a function of the axial quadrupole moment, and it appears that $\langle \hat{Q}_{20}\rangle\simeq 15$~b corresponds to the ground-state deformation.

\begin{figure}
\includegraphics[width=8.5cm]{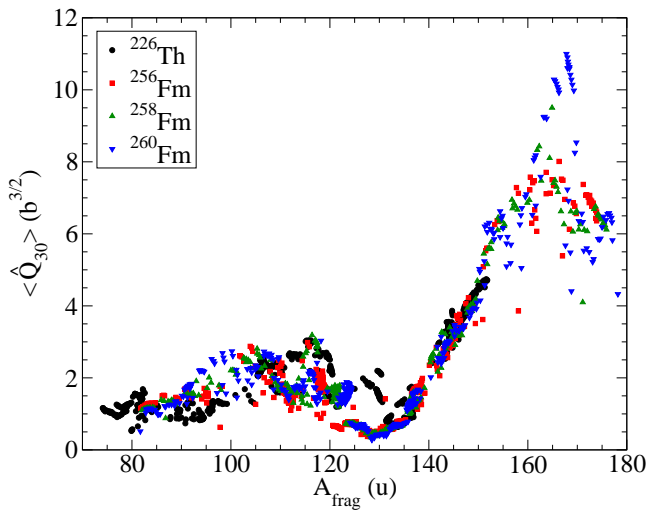}
  \caption{(Color online) Axial mass octupole moments $\langle\hat{Q}_{30}\rangle$ of nascent fission fragments for ${}^{226}$Th and ${}^{256-260}$Fm. \label{fragdefq3}}
\end{figure}

Axial mass octupole moments of fission fragments are plotted in Fig.~\ref{fragdefq3} as functions of the fragment mass. The octupole moments display almost the same behavior versus $A_{\text{frag}}$ as the one for the quadrupole moments: minima are observed for $A\simeq 86$ and $A\simeq 130$ and maxima for $A\simeq 112$ and $A\simeq 170$.

\subsection{Fragment deformation energy}

Energy partitioning in fission is a key input of models aiming at describing sequential neutron and $\gamma$-ray emission from fission fragments \cite{lemaire05,lemaire06,kornilov07}. In the present study, the assumption will be made that the excitation energy stored into fission fragment arises only from their quadrupole and octupole deformations at the moment of scission. With this assumption possible intrinsic or thermal excitations prior to scission are neglected. The estimates given below must therefore be considered as lower bound of fragment excitation energies.

The fragment deformation energy is defined as~\cite{simonphd}
\begin{equation}
E_{\text{def}}= E_{\text{ff}}-E_{\text{gs}},
\end{equation}
where $E_{\text{ff}}$ is the energy of the nascent fragment, and $E_{\text{gs}}$ the one of the fragment  ground state. In this work, $E_{\text{gs}}$ has been deduced for all fragments from usual HFB calculations, whereas $E_{\text{ff}}$ is the HFB energy predicted in a constrained HFB calculation where the axial quadrupole and octupole moments are those obtained at scission configurations (see Figs. \ref{fragdefq2} and \ref{fragdefq3}). Let us mention that in these two sets of calculations, the neutron and proton numbers of each fragment have been taken to be integer values closest to the N and Z mean values calculated for the nascent fragments. Such an approximation leads to an uncertainty in HFB energies which amounts to be less than 1 MeV.

\begin{figure}
\includegraphics[width=8.5cm]{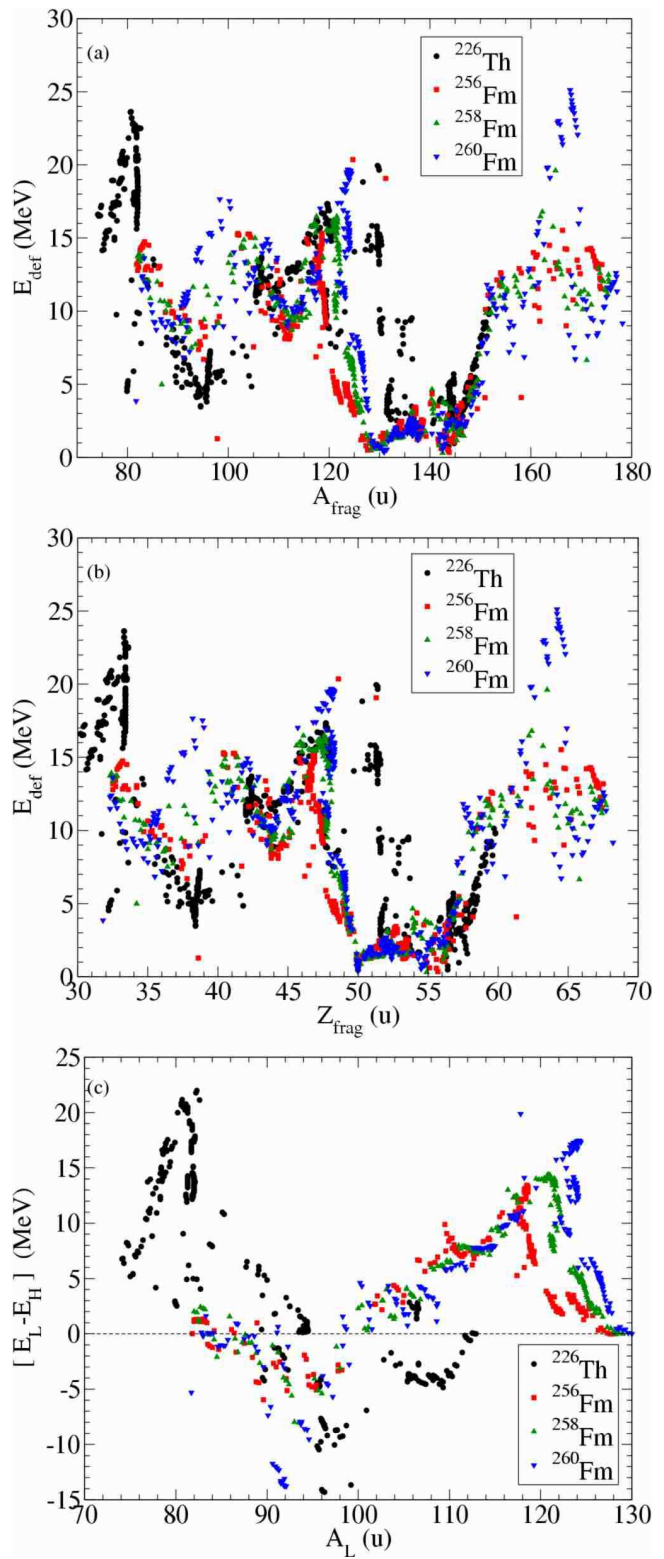}
  \caption{(Color online) Nascent fragment deformation energies for ${}^{226}$Th and ${}^{256-260}$Fm as functions of: (a) fragment mass, and (b) fragment charge. The differences between light (L) and heavy (H) fragment energies as functions of light fragment mass are shown in (c).\label{enpartition}}
\end{figure}

The FF deformation energies ($E_{\text{def}}$) derived in this way for the four nuclei studied here are shown as functions of $A_{\text{frag}}$ and $Z_{\text{frag}}$ in Figs.~\ref{enpartition}(a) and \ref{enpartition}(b), respectively. Strong variations are observed, with maxima reaching $E_{\text{def}}\sim 15-20$~MeV near $A_{\text{frag}}\sim 120$ and minima close to zero near $A_{\text{frag}}\sim 145$ and $A_{\text{frag}}\sim 130$. This latter minimum corresponds to symmetric division in Fm nuclides. Its origin is of course the occurrence of strong shell effects in nuclei close to ${}^{132}$Sn.

When plotted as function of $Z_{\text{frag}}$, regions with $E_{\text{def}}\sim 0$ correspond to $Z_{\text{frag}}\sim 50$ and $Z_{\text{frag}}\sim 56$, that is to ${}^{128-130}$Sn and ${}^{144}$Ba, respectively. Furthermore, the maximum identified previously in the $E_{\text{def}}$ values at $A_{\text{frag}}\sim 120$ gets split over two $Z_{\text{frag}}$ components, namely $Z_{\text{frag}}\sim 48$ and $Z_{\text{frag}}\sim 52$, that is for near symmetric and highly asymmetric charge divisions in the Fm and Th nuclides, respectively.

Finally, Fig.~\ref{enpartition}(c) displays the difference ($E_{\text{L}}-E_{\text{H}}$) between the deformation energies of light and heavy fragments. This difference takes on values ranging from 23~MeV to -15~MeV. Extrema are located at far-asymmetric mass divisions. More than 70\% of the light fragments display $E_{\text{L}}>E_{\text{H}}$ values.

\subsection{Prompt fission neutrons}

In the present section, we aim at calculating the multiplicity $\nu_{\text{frag}}$ of prompt neutrons emitted by each fission fragment. For this purpose, we assume that the deformation energy of any fragment is converted into internal excitation energy through collective vibrations and that the fragment will de-excite only through prompt neutron emission. As an estimate, the neutron emission multiplicity of one fragment is taken as~\cite{ruben90,knitter91}
\begin{equation}
\nu_{\text{frag}}=\frac{E_{\text{def}}}{\langle E_k\rangle + 
B_{\text{n}}^*}\text{ },\label{nubarexpr}
\end{equation}
where $B_n^*$ is the one-neutron binding energy in nascent fragment, and $\langle E_k \rangle$ the mean energy of the emitted neutron. The latter is assumed to be 2~MeV in ${}^{226}$Th \cite{madland06} and 1.5~MeV in ${}^{256-260}$Fm \cite{budtz88}.

\subsubsection{One-neutron binding energy}

\begin{figure}
\includegraphics[width=8.5cm]{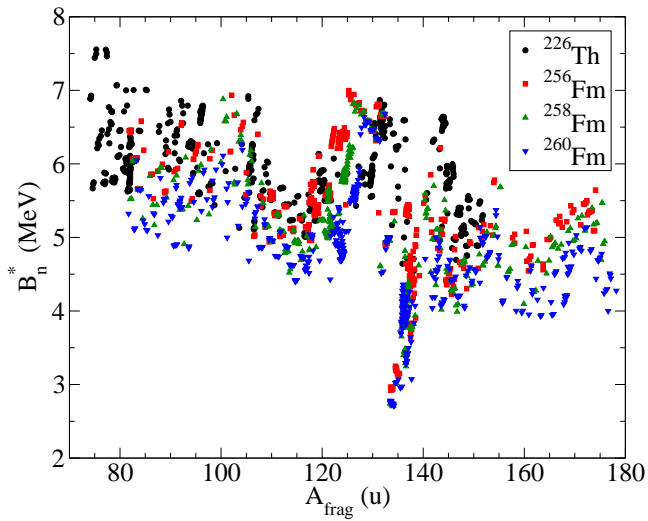}
  \caption{(Color online) One-neutron binding energies of nascent fission fragments as functions of fragment mass for ${}^{226}Th$ and ${}^{256-260}$Fm\label{refbn}.}
\end{figure}

In the present work the one-neutron FF binding energy $B_{\text{n}}^*$ is taken equal to the neutron chemical potential obtained in the HFB calculations performed on the scission line. The $B_{\text{n}}^*$ values are plotted in Fig.~\ref{refbn} as a function of fragment mass. It is seen that they globally decrease from approximately 7~MeV to 3~-~4~MeV with increasing mass. The lowest $B_{\text{n}}^*$ values are obtained for $A\simeq 136$ ($Z\simeq 50$, $N\simeq 86$). 

\begin{figure}
\includegraphics[width=8.5cm]{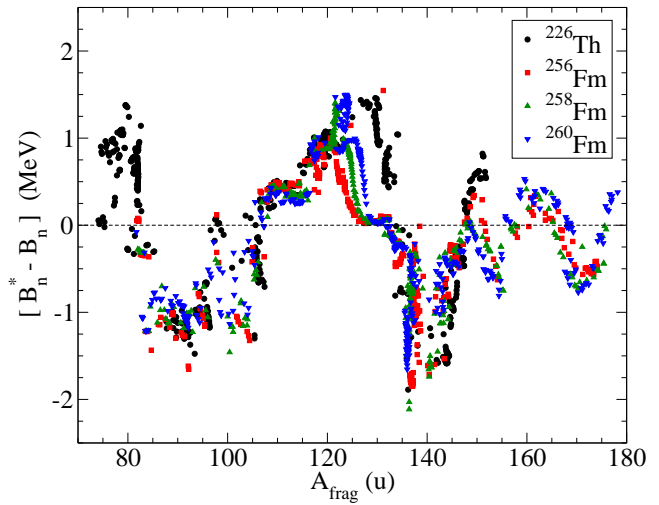}
  \caption{(Color online) Differences between one-neutron binding energies of fragments as calculated for scission configurations and for ground states, plotted as functions of fragment mass.\label{diffbnvsaf}}
\end{figure}

The difference between the calculated one-neutron binding energies of the fragment at scission $B_{\text{n}}^*$ and ground state $B_{\text{n}}$ is plotted in Fig.~\ref{diffbnvsaf}. We find that this difference can be as large as 2 MeV in absolute value. Such differences are a consequence of the evolution of single-particle neutron gaps as a function of deformation.

\subsubsection{Neutron multiplicity}

\begin{figure}
\includegraphics[width=8.5cm,angle=0]{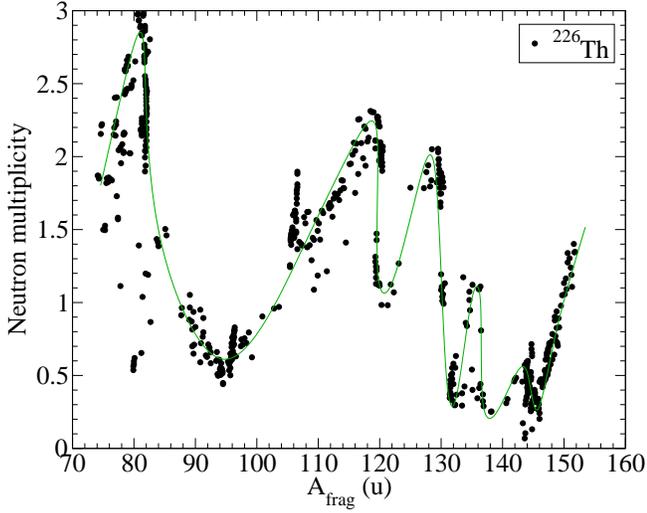}
\caption{(Color online) ${}^{226}$Th. Calculated neutron multiplicity as a function of fragment mass. The solid line is to guide the eye. \label{nubar226th}}
\end{figure}

\begin{figure}
\includegraphics[width=8.5cm,angle=0]{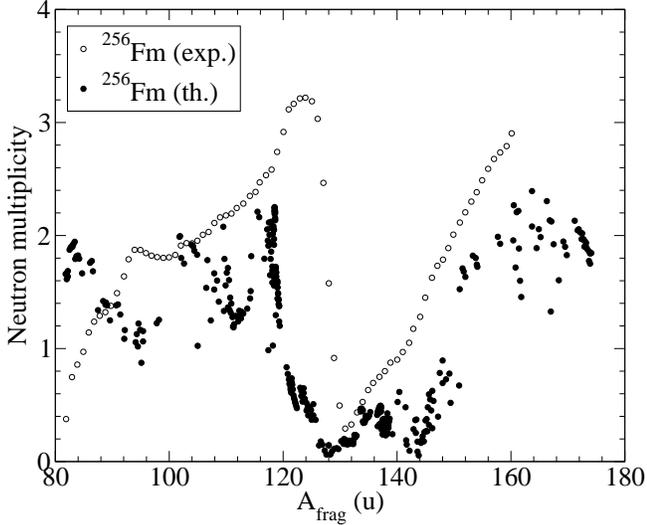}
\caption{${}^{256}$Fm. Neutron multiplicity versus fragment mass. Comparison between predictions (solid symbols) and data~\cite{gindler79} (empty symbols).\label{nubarexp}}
\end{figure}

\begin{figure}
\includegraphics[width=8.5cm,angle=0]{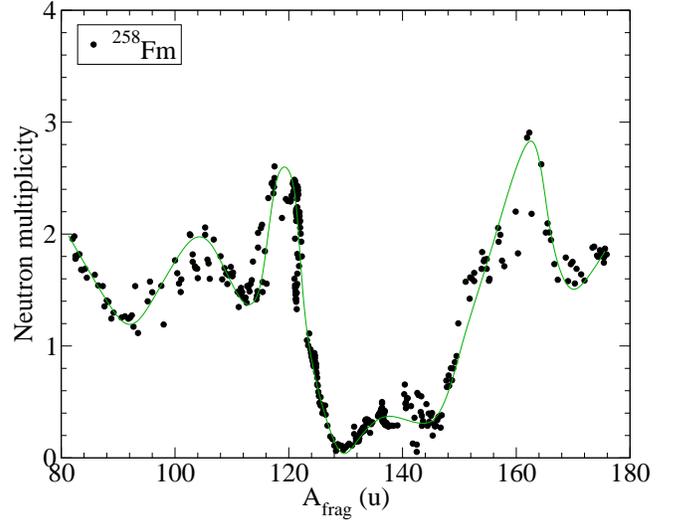}
\includegraphics[width=8.5cm,angle=0]{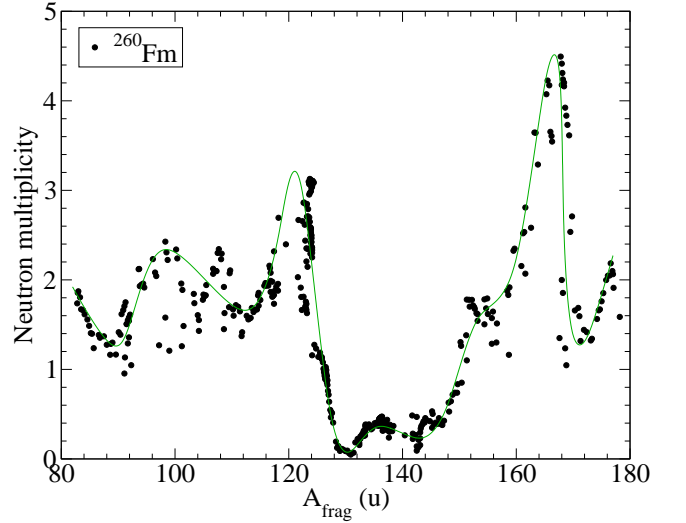}
\caption{(Color online) Neutron multiplicities for ${}^{258}$Fm (upper panel) and ${}^{260}$Fm (lower panel). Lines are to guide the eye.\label{nubar260fm}}
\end{figure}

The multiplicities calculated from Eq.~\eqref{nubarexpr} are shown in Figs.~\ref{nubar226th}~-~\ref{nubar260fm} as functions of fragment mass for the four studied nuclei. On figures \ref{nubar226th} and \ref{nubar260fm}, a solid line has been added to guide the eye. Typical saw-tooth structures are observed, displaying maxima and minima. These structures appear correlated with the quadrupole deformation of fragments at scission.

In the case of ${}^{226}$Th fission, the neutron multiplicity curve displays pronounced structures  separated by five mass units from $A_{\text{frag}}=110$ to $A_{\text{frag}}=150$, that are linked to: i) fragment deformations (Figs.~\ref{fragdefq2} and \ref{fragdefq3}), ii) the deformability of the fragments  (the softness of potential energies with respect to axial quadrupole and octupole deformations), and iii) the one-neutron binding energy (see Fig.~\ref{refbn}).

For Fm isotopes, the curves look more regular, and show that neutron emission is almost vanishing around  $A_{\text{frag}}=130$ and is maximum around $A_{\text{frag}}=120$. In Fig.~\ref{nubarexp}, comparison is made with the experimental data for spontaneous fission of ${}^{256}$Fm \cite{gindler79}. The agreement between theoretical values and measurements is rather satisfactory, as the global data pattern is well reproduced. However, calculations appear to underestimate the number of emitted neutrons in the $A_{\text{frag}}=90-130$ region and a second minimum is found around $A_{\text{frag}}=144$. As a consequence, the calculated number of emitted neutrons is 30\% smaller than experimental data.


These discrepancies probably come from our model assumptions. Part of the overall underestimation of the number of emitted neutrons is presumably due to the fact that the deformation energy has been calculated with constraints placed only on axial quadrupole and octupole deformations of fragments. More realistic calculations should include the effect of higher order multipole fragment deformations such as $q_{40}$ and $q_{60}$. As for the second minimum at $A_{\text{frag}}\sim 144$, we think it may originate from the method adopted to define the two fragments, especially for cases where many particles are present in the neck. Handling the neck rupture in a more realistic manner and including higher order multipole deformations in the calculation of deformation energies would change our predictions. However we have doubt that these considerations, alone, would be pertinent enough for bringing in significant improvements. These discussion will be further extended below.

\subsection{Deviation from the unchanged charge distribution}

Introduced in 1962 by Wahl, the fragment unchanged charge distribution $Z_{\text{UCD}}$ (i.e. charge polarization) is the charge number of a fragment with a given mass $A_{\text{frag}}$, if its $Z/A$ ratio were the same as the one of the fissioning nucleus \cite{wahl62}:
\begin{equation}
Z_{\text{UCD}}\equiv \frac{Z_{\text{fs}}.A_{\text{frag}}}{A_{\text{fs}}}.
\end{equation}

\begin{figure}
\includegraphics[width=8.5cm]{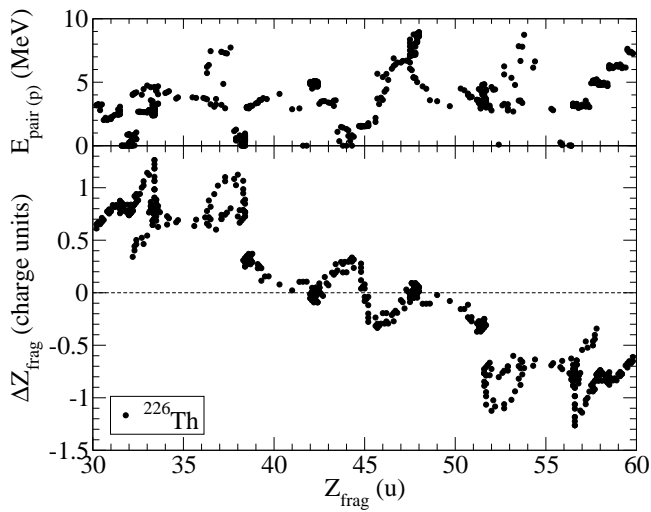}
  \caption{Nascent fission fragment proton pairing energy (upper panel) and deviation from unchanged charge distribution (bottom panel) as functions of fragment charge for ${}^{226}$Th. \label{pairucd226Th}}
\end{figure}

\begin{figure}
\includegraphics[width=8.5cm]{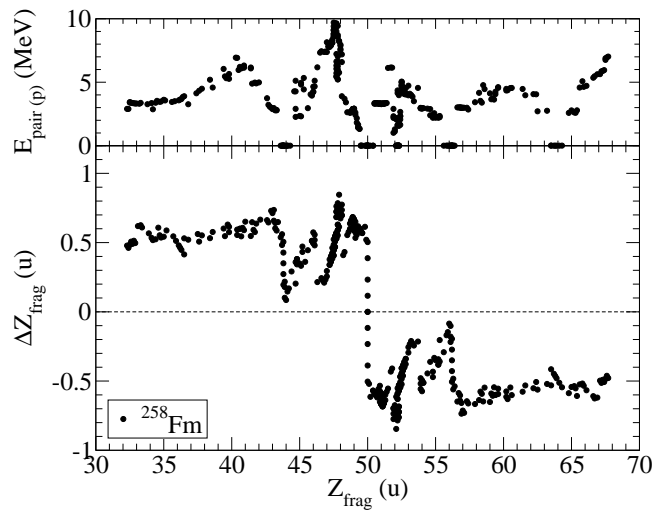}
  \caption{Same as Fig. \ref{pairucd226Th} for ${}^{258}$Fm.\label{pairucd258Fm}}
\end{figure}

The deviation $\Delta Z_{\text{frag}}=Z_{\text{frag}}-Z_{\text{UCD}}$ of the charge of the fragment $Z_{\text{frag}}$ from the unchanged charge distribution $Z_{\text{UCD}}$ is plotted in Figs.~\ref{pairucd226Th} and \ref{pairucd258Fm}  as a function of the fragment charge for ${}^{226}$Th and ${}^{258}$Fm. Values of $\Delta Z_{\text{frag}}$ for ${}^{256}$Fm and ${}^{260}$Fm are very close to the ones of ${}^{258}$Fm. We first observe that $\Delta Z_{\text{frag}}$ is globally positive for light fragments and negative for heavy ones. This feature stems from the fact that heavy systems may sustain stronger neutron excess than light ones, as observed and discussed for several fissioning systems~\cite{bocquet89,gonnenwein91}. The patterns displayed by $\Delta Z_{\text{frag}}$ as functions of $Z_{\text{frag}}$ are quite contrasted. While both sets of $\Delta Z_{\text{frag}}$ values show sharp structures as $Z_{\text{frag}}$ increases, $\Delta Z_{\text{frag}}$ globally decreases in ${}^{226}$Th from $\Delta Z_{\text{frag}}\simeq 1$ to $\Delta Z_{\text{frag}}\simeq -1$. In contrast, in ${}^{258}$Fm the $\Delta Z_{\text{frag}}$ values reach a plateau with $|\Delta Z_{\text{frag}}|\simeq 0.5$ for $Z_\text{frag}> 57$ and $Z_\text{frag}< 43$. In an attempt to understand the origins of these sharp structures and different global trends, we have sought for possible correlations with other structure properties, namely proton separation energies of: i) fissioning nuclei along scission lines, and ii) each nascent fragment. No clear cut correlation is found. However the structures observed in the $\Delta Z_{\text{frag}}$ values for both nuclides seem to coincide with the variations of the fragment pairing energies $E_{\text{pair (p)}}$ as can be seen when comparing the plots in the upper and  bottom panels in Figs.~\ref{pairucd226Th} and~\ref{pairucd258Fm}.


\subsection{Total kinetic energy}
\subsubsection{Distance between fragments}
\begin{figure}
\includegraphics[width=8.5cm]{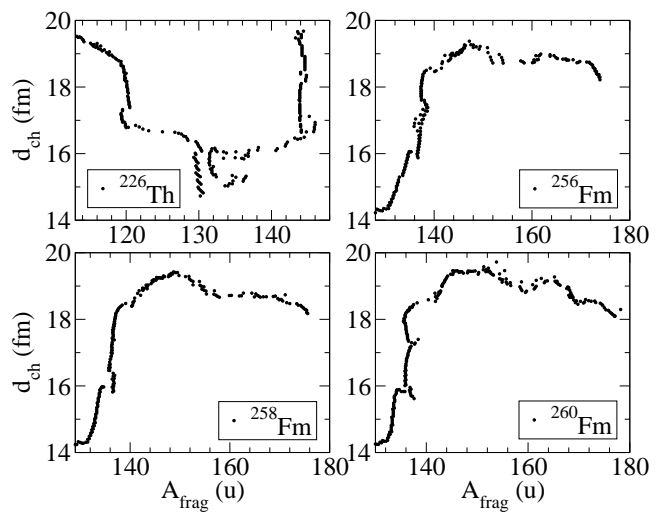}
  \caption{Distances between nascent fragment centers of charge calculated as functions of fragment mass for ${}^{226}$Th and ${}^{256-260}$Fm.\label{dists}}
\end{figure}

The total kinetic energy (TKE) of a given fragmentation can be estimated from the formula
\begin{equation}
E_{\text{TKE}}=\frac{e^2Z_{\text{H}}Z_{\text{L}}}{d_{\text{ch}}}\label{eqtke},
\end{equation}
where $e$ is the electron charge, $Z_{\text{H}}$ ($Z_{\text{L}}$) the charge of the heavy (light) fragment, and $d_{\text{ch}}$ the distance between fragment centers of charge at scission. The distance $d_{\text{ch}}$ deduced from our calculations is plotted as a function of fragment mass for ${}^{226}$Th, ${}^{256}$Fm, ${}^{258}$Fm and ${}^{260}$Fm in Fig.~\ref{dists}. For all considered nuclei, the distance between fragment centers of charge at scission falls in the range $d_{\text{ch}}=14-20$~fm. The distance $d_{\text{cm}}$ between centers of mass has also been calculated. The difference $\delta d$ between $d_{\text{cm}}$ and $d_{\text{ch}}$ appear rather small: $\delta d\simeq 0.08$~fm in ${}^{226}$Th and $\delta d\simeq 0.05$~fm in ${}^{256-260}$Fm.

\subsubsection{Total kinetic energy for ${}^{226}$Th}

\begin{figure}
\includegraphics[width=8.5cm]{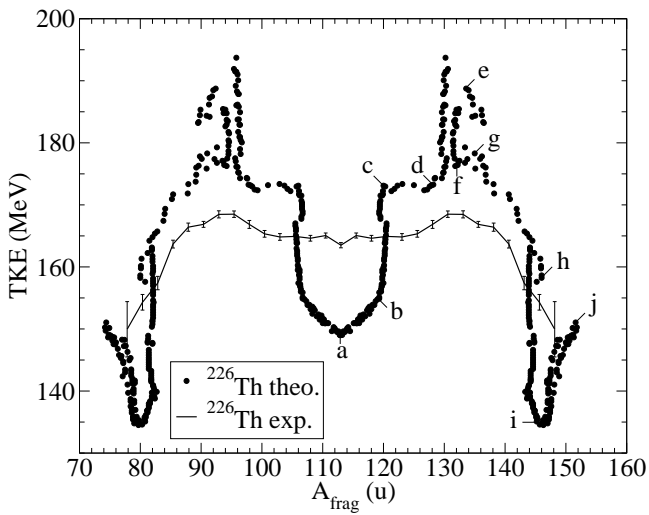}
  \caption{${}^{226}$Th. TKE values of nascent fission fragments as a function of fragment mass. Comparison between predictions (solid dots) and data~\cite{schmidt01b,bockstiegel97}.\label{exptketh}}
\end{figure}

The TKE values of ${}^{226}$Th fission fragments are plotted as functions of fragment mass in Fig.~\ref{exptketh}. The dots represent the result of Eq.~\eqref{eqtke} whereas the solid line follows the experimental data of Ref.~\cite{bockstiegel97,schmidt01b} obtained in electro-magnetic induced fission measurements. One notices that theoretical results present many more structures than do experimental data. This difference may be explained from the fact that the experimental measurements correspond to an excitation energy of the fissioning nucleus of the order of 11~MeV, whereas formula~\eqref{eqtke} is valid only for low energy fission. As well known, an increase in the fission energy smooths out kinetic energy distribution. In particular the kinetic energy in the symmetric mass region increases \cite{pomme94} which explains why experimental TKE display only a very shallow minimum for $A_{\text{frag}}=A/2$.

The experimental TKE values display maxima for $A_{\text{frag}}\simeq 132$ ($Z_{\text{frag}}\simeq 54$) and  $A_{\text{frag}}\simeq 94$ ($Z_{\text{frag}}\simeq 36$), whose positions coincide with those  found in our calculations. The sharp structures in the theoretical TKE stem from the compactness of scission configurations marked with the symbols d, e and f in Fig.~\ref{scission_lines}. They do not show up in the experimental TKE values, as details of the energy landscape along the scission line are presumably washed out by the dynamics of the Coulomb induced fission  process~\cite{schmidt01a}. Nonetheless, theoretical results are in qualitative agreement with experimental data, with deviations never exceeding 15\%, and the calculated mean TKE value $(\overline{\text{TKE}})_{\text{th}}\sim 169$~MeV, is found close to the experimental mean value $(\overline{\text{TKE}})_{\text{exp}}=167.7\pm3.4$~MeV~\cite{schmidt00}. 

\subsubsection{Total kinetic energy for ${}^{256-260}$Fm.}

\begin{figure}
\includegraphics[width=8.5cm]{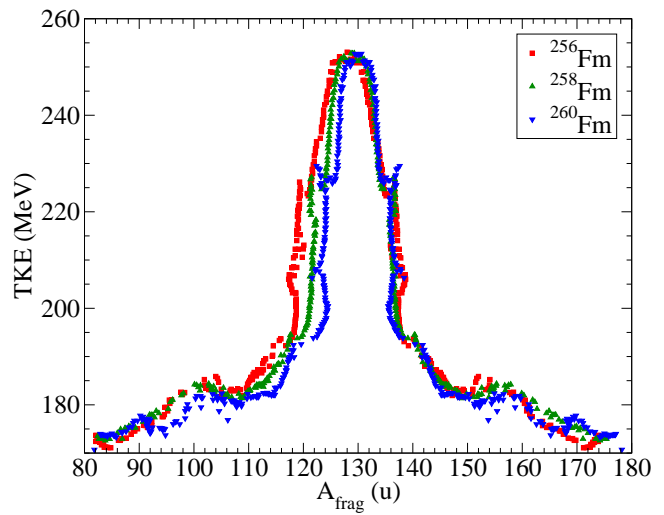}
\caption{(Color online) ${}^{256-260}$Fm. TKE values of nascent fission fragments as functions of fragment mass.\label{tkeallfm}}
\end{figure}

The TKE values of ${}^{256-260}$Fm fission fragments are plotted in Fig.~\ref{tkeallfm} as functions of fragment mass. The TKE curves look rather similar in all three isotopes. They display a sharp peak reaching $\text{TKE}\simeq 250$~MeV for symmetric fission. These features are characteristic of compact scission, where the fissioning system gives rise to nearly spherical fragments separated by a small distance. Furthermore, we also observe that the full width $\Delta_{\text{TKE}}$ at half maximum is narrowing from $\Delta_{\text{TKE}}=20$~u to $\Delta_{\text{TKE}}=14$~u in going from ${}^{256}$Fm to ${}^{260}$Fm.

\begin{figure}
\includegraphics[width=8.5cm]{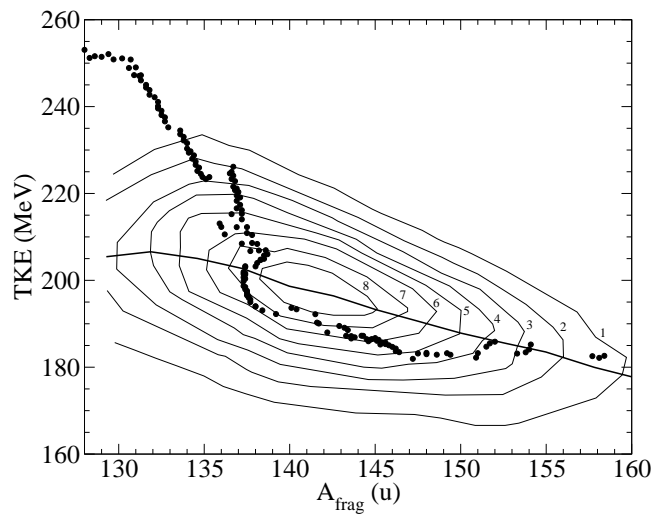}
  \caption{${}^{256}$Fm. TKE values of fission fragments as functions of fragment mass. Results of theoretical calculations (dots) are displayed with pre-neutron-emission data (contour diagram). The solid line represents the experimental average TKE~\cite{hoffman80a}. \label{exptkefm}}
\end{figure}
Theoretical TKE in ${}^{256}$Fm (black points) are compared with experimental ones~\cite{hoffman80a} in Fig.~\ref{exptkefm}. Calculated results are found in very good agreement with the average TKE data for asymmetric fission ($A_{\text{frag}}=138-150$). However, they overestimate the experimental values for $A_{\text{frag}}< 138$ by up to $16\%$. This latter feature may indicate that the calculated distance between fragment charges $d_{\rm ch}$ is too small for scission configurations close to symmetry. This remark is consistent with the fact that for the same fragmentations, $\nu_{\rm frag}$ values calculated from deformation energies is underestimated (see Fig.~\ref{nubarexp}). These under- and over-estimations are interpreted as due to our study, restricted to the ($q_{20}$,$q_{30}$) deformations, which favors compact scission. Accessing elongated fission configurations for nearly-symmetric fragmentations of ${}^{256}$Fm implies that at least three collective coordinates must be considered in constrained HFB calculations.

\section{Conclusion}

In this work, large scale HFB calculations using the Gogny D1S force have been performed in order to investigate structure properties of ${}^{226}$Th and ${}^{256-260}$Fm at scission and the characteristics of fission fragments along scission lines. Scission configurations are first analyzed assuming that axial quadrupole and octupole collective coordinates play a major role in fission. We have found from our constrained HFB calculations that the scission mechanism depends on which heavy nuclide and fragmentation are considered. This mechanism may display either a smooth or an abrupt character in the ($q_{20}$,$q_{30}$) plane. The former property means that the potential energy of the fissioning system changes smoothly over deformation from outer saddle to scission and beyond where Coulomb repulsion takes place between fission fragments. This scission property is found for the Fm symmetric fragmentations. For asymmetric fragmentations in all nuclei of present interest, there is a sudden drop in potential energies whenever scission takes place. To accommodate with these contrasted properties, post-scission points are defined for matter densities present in the neck that are weaker than $\rho=0.06$~fm${}^{-3}$. With this criterion, scission lines are calculated, and the fragmentations determined assuming sharp cuts across the necks.

Properties of fission fragments and correlations with properties of fissioning systems along scission lines have been discussed. These comprise potential and deformation energies, quadrupole and octupole deformations, total kinetic energies, prompt neutron emissions, deviation from unchanged charge distribution, and energy partitioning. All these properties reflect either shell and/or pairing contents of potential energies of both fission fragments and fissioning nuclei, in particular for multipole deformations, neutron multiplicities, and total kinetic energies. Predictions are found in reasonably good agreement with experimental data for total kinetic energy (${}^{226}$Th, ${}^{256}$Fm) and prompt neutron multiplicity (${}^{256}$Fm) of fission fragments.

The present microscopic analysis shows that the structure of the two-dimensional ($q_{20}$, $q_{30}$) potential energy surface in ${}^{226}$Th is similar to those previously calculated in U and Pu. The different behavior with respect to scission found in Fm isotopes and the fact that symmetric elongated configurations do not appear in our description may indicate that a collective space with more that two dimensions is needed to describe scission configurations and fragment properties in these nuclei. Preliminary investigations show that other heavy actinides probably also require an enlargement of the dimension of the collective space used. In view of the encouraging results obtained so far, in particular in light actinides, it seems worth attempting to extend the present static calculation to three dimensions or even more. Of course, a description of fission observables such as fragment mass and kinetic energy distributions will also require to extend the two-dimensional dynamical model employed in Uranium~\cite{goutte05} to higher dimensions.

\section*{ACKNOWLEDGMENTS}

We gratefully acknowledge D.~Gogny and J.-F.~Berger for stimulating and enlightening discussions.

\bibliography{dubray}

\end{document}